\documentclass{PoS}
\usepackage{amsmath}

\title{The effect of neutrino quantum decoherence}

\ShortTitle{The effect of neutrino quantum decoherence}

\author{\speaker{Konstantin Stankevich} \\ 
        Department of Theoretical Physics, Lomonosov Moscow State University, 119992 Moscow, Russia\\
        E-mail: \email{kl.stankevich@physics.msu.ru}}

\author{Alexander Studenikin\\
        Department of Theoretical Physics, Lomonosov Moscow State University, 119992 Moscow, Russia\\
        Joint Institute for Nuclear Research, 141980 Dubna, Moscow Region, Russia\\
        E-mail: \email{studenik@srd.sinp.msu.ru}}

\abstract{The neutrino oscillation patterns can be modified by neutrino interactions with external environments including electromagnetic fields that can influence on neutrinos in the case neutrinos have nonzero electromagnetic properties \cite{Giunti_Studenik}. The phenomenon of neutrino oscillations can proceed only in the case of the coherent superposition of neutrino mass states. An external environment can modify a neutrino evolution in a way that conditions for the coherent superposition of neutrino mass states are violated. Such a violation is called quantum decoherence of neutrino states and leads to the suppression of flavor neutrino oscillations \cite{Stankevich_Studenikin,Stankevich_Studenikin_2}. Note that neutrino decoherence appeared due to the wave separation of different mass states is usually not related to quantum neutrino decoherence, the effect that is not considered below. We consider the neutrino quantum decoherence due to neutrino radiative decay in the presence of an electron medium and radiation field. The corresponding damping of neutrino oscillations is calculated. 
}

\FullConference{
European Physical Society Conference on High Energy Physics - EPS-HEP2019 -\\
			10-17 July, 2019\\
			Ghent, Belgium}

\begin{document}


The quantum neutrino decoherence has attracted a growing interest during the last 15 years. Within reasonable amount of the performed studies the method based on the Lindblad equation \cite{Lindblad, Gorini_Kossakowski} for describing neutrino evolution has been used. This approach is the most general and gives the possibility to study neutrino quantum  decoherence as a consequence of interactions of a neutrino system with matter, nonstandard interactions, quantum foam and quantum gravity (see, for example, \cite{Oliveira2016,Joao_Coelho2}). Nevertheless, in the Lindblad approach the quantum decoherence is described by the free parameter that can be found only from the experimental data. An alternative way to study neutrino quantum decoherence is based on summing over all degrees of freedom of an external environment. On this basis, for instance in \cite{Burgess_Michaud, Benatti_Florianini_2}, the influence of matter fluctuations is considered.

In \cite{Stankevich_Studenikin,Stankevich_Studenikin_2} we suggested the new theoretical approach for studying of neutrino quantum decoherence that is based on formalism of quantum electrodynamics of open systems \cite{Breuer_Pettrucione}. This new approach gives the possibility to find the exact form of the dissipative term and the decoherence parameter. It is shown that the studied phenomenon can be significant for description of neutrino oscillations in extreme conditions of astrophysical environments peculiar to supernovae, neutron stars or quasars.


For description of the neutrino decoherence we use the formalism of quantum electrodynamics of open systems which was used in \cite{Breuer_Pettrucione} for evolution of electrons. We start with the quantum Liouville equation for the density matrix of a system composed of neutrinos and an electromagnetic field

\begin{equation}
\frac{\partial}{\partial t} \rho =  - i \int d^3 x\left[ H(x),\rho \right]
\label{rho}
,
\end{equation}
where $H(x) = H_\nu(x) + H_f(x) +H_{int}(x)$ is the Hamiltonian density of the system, $H_\nu(x)$ and $H_f(x)$ are the Hamiltonian densities of the neutrino system and the electromagnetic field respectively, and $H_{int} (x) = j_\alpha (x) A^\alpha (x)$ describes interaction between neutrino and the electromagnetic field, where $j_\alpha (x)$ is the current density of neutrino and $A_\alpha$ is the electromagnetic field. Here below we propose and study a new mechanism of the quantum decoherence that appears due to the neutrino radiative decay. In this case the current density can be expressed in the following form \cite{Olivio_Nieves_Pal}

\begin{equation}
j_\alpha (x) = \overline{\nu}_i(x)\Gamma_\alpha \nu_j(x)
\label{current0}
,
\end{equation}
where $\nu_i(x)$ is the neutrino field with mass $m_i$ and $\Gamma_\alpha$ is an effective electromagnetic vertex $\Gamma_\alpha = U^*_{ei} U_{ej} \tau_{\alpha\beta} \gamma^\alpha L,$
where $U$ is the lepton mixing matrix and $L = \frac 1 2 (1-\gamma_5)$ is the projection operator for the left-handed fermions. It is supposed that the four-velocity of the center of mass of the electron background is at rest. In the case of a nonrelativistic (NR) background $\tau_{\alpha\beta}$ can be expressed as \cite{Olivio_Nieves_Pal}

\begin{equation}
\tau_{\alpha\beta}^{NR} =\tau^{NR} P_{\alpha\beta}  = - \sqrt{2} \frac{e G_F n_e} {m_e} P_{\alpha\beta}
.
\end{equation}
In the extreme relativistic (ER) case, when the temperature of the background electrons $T\gg m_e$, one gets

\begin{equation}
\tau_{\alpha\beta}^{ER} =\tau^{ER} P_{\alpha\beta} = - \frac{e G_F T^2}{2 \sqrt{2}} P_{\alpha\beta}
\label{ER}
.
\end{equation}
For the degenerate electron gas of neutron stars we get

\begin{equation}
\tau_{\alpha\beta}^{Deg} =\tau^{Deg} P_{\alpha\beta} = - \dfrac{\sqrt 2 e G_F}{4} \left(\dfrac{3 n_e}{\pi}\right)^{2/3} P_{\alpha\beta}
.
\end{equation}
The tensor $P_{\alpha\beta} = \delta_{\alpha\beta} - \frac{k_\alpha k_\beta}{|\vec k|^2}$ is a projector onto the transverse component in k-space, $e$, $m_e$ and $n_e$ are the charge and mass of an electron and the medium number density, respectively.

Using the analogous calculations to those performed in \cite{Fabbricatore_Grigoriev,Pustoshn} we express the current density (\ref{current0}) in the form

\begin{equation}
j_3 = 2 U^*_{ei} U^*_{ej} \tau
\left(
\begin{matrix}
0 & 1 \\
1 & 0
\end{matrix}
\right)
,
\label{current}
\end{equation}
where for different cases $\tau$ stands for  $\tau^{ER}$, $\tau^{NR}$ or $\tau^{deg}$. It is convenient to decompose the current (\ref{current}) on the eigenoperators of the neutrino Hamiltonian

\begin{equation}
j_{\pm} = 2 U^*_{ei} U^*_{ej} \tau \sigma_{\pm}
\label{operators}
,\ \ \ 
\sigma_+ =
\left(
\begin{matrix}
0 & 1 \\
0 & 0
\end{matrix}
\right),
\sigma_- =
\left(
\begin{matrix}
0 & 0 \\
1 & 0
\end{matrix}
\right)
.
\end{equation}

The equation (\ref{rho}) can be formally solved (integrated). Since we are not interested in the evolution of the electromagnetic field, its degrees of freedom should be traced out

\begin{equation}
\rho_\nu(t_f) = tr_f \left( Texp\left[ \int_{t_i}^{t_f} d^4x  \left[ H(x),\rho(t) \right]\right] \right)
,
\end{equation}
where $\rho_\nu(t) = tr_f \rho(t)$ is the density matrix which describes the evolution of the neutrino system. 

In the second-order approximation we get the quantum optical master equation for the neutrino case, which is analogous to one for the case of electrons \cite{Breuer_Pettrucione},

\begin{equation}
\frac{\partial}{\partial t} \rho_\nu (t) = - i \left[ H_\nu,\rho_\nu(t) \right] - i \left[ H_{S},\rho_\nu(t) \right]+D(\rho_\nu(t))
.
\label{OpticEquation}
\end{equation}

The Hamiltonian $H_S$ leads to a renormalization of the system Hamiltonian $H_\nu$ which is induced by the vacuum fluctuations of the radiation field and by thermally induced processes. Our goal is to find the dissipative terms, thus we omit the renormalization  part $H_S$ in the following derivations. In (\ref{OpticEquation}) $D(\rho_\nu(t))$ is a dissipator of the equation which can be expressed in the following form (see \cite{Breuer_Pettrucione})

\begin{multline}
D(\rho_\nu(t)) = \frac{\Delta_{ij}}{4 \pi^2}(f(2\Delta_{ij})+1) \left( j_-\rho_\nu (t) j_+ - \frac{1}{2} j_+j_-\rho_\nu(t) - \frac{1}{2} \rho_\nu(t) j_+j_-     \right)+\\
+\frac{\Delta_{ij}}{4 \pi^2} f(2\Delta_{ij}) \left( j_+\rho_\nu (t) j_- - \frac{1}{2} j_-j_+\rho_\nu(t) - \frac{1}{2} \rho_\nu(t) j_-j_+ \right)
,
\label{dissipator}
\end{multline}
where $\Delta_{ij}$ is the energy difference between two neutrino mass states $\nu_i$ and $\nu_j$, and $f(E)$ denotes the Planck distribution $f(E) = \frac{1}{e^{E/kT}-1}.$ The first term in equation (\ref{dissipator}) is responsible for the spontaneous and the thermally induced emission process and the second one is responsible for the thermally induced absorption process.

In the medium, it is necessary to define new neutrino effective mass states $\tilde{\nu}$ and the effective mixing angle $\tilde{\theta}$. In this new basis the neutrino evolution Hamiltonian is diagonal (see, for example \cite{Pal} and \cite{Freund})  and the energy difference between two neutrino states is expressed as

\begin{equation}
\Delta_{ij} = \dfrac{\sqrt{(\Delta m_{ij} \cos 2\theta_{ij} - A)^2 +\Delta m^2_{ij} \sin^2 2 \theta_{ij}}}{2 E}
,
\end{equation}
where the effective mixing angle is given by

\begin{equation}
\sin^2 2 \tilde{\theta}_{ij} = \dfrac{\Delta m^2_{ij} \sin^2 2 \theta_{ij}}{(\Delta m_{ij} \cos 2 \theta_{ij} - A)^2 + \Delta m_{ij}^2 \sin^2 2 \theta_{ij}}
,
\end{equation}
where  $A = 2 \sqrt{2} G_F n_e E$ and $E$ is the neutrino energy. Putting everything together we get the final expression for neutrino evolution in the effective mass basis

\begin{multline}
\frac{\partial}{\partial t} \rho_{\tilde{\nu}} (t) = - i \left[ H_{\tilde{\nu}},\rho_{\tilde{\nu}}(t) \right] +  \\
+\kappa_1 \left( \sigma_-\rho_{\tilde{\nu}} (t) \sigma_+ - \frac{1}{2} \sigma_+\sigma_-\rho_{\tilde{\nu}}(t) - \frac{1}{2} \rho_{\tilde{\nu}}(t) \sigma_+\sigma_-     \right)+\\
+\kappa_2 \left( \sigma_+\rho_{\tilde{\nu}} (t) \sigma_- - \frac{1}{2} \sigma_-\sigma_+\rho_{\tilde{\nu}}(t) - \frac{1}{2} \rho_{\tilde{\nu}}(t) \sigma_-\sigma_+ \right)
,
\label{Optic2}
\end{multline}
where the Hamiltonian $H_{\tilde{\nu}} = diag(\tilde{E_1},\tilde{E_2})$ and $\kappa_1$ and $\kappa_2$ are the decoherence parameters of the neutrino system. For extreme environment $\kappa_1 \approx \kappa_2 = \kappa$

\begin{equation}
\kappa = \dfrac{\Delta_{ij}}{\pi^2} \sin^2 2\tilde{\theta}_{ij} \tau^2 f(2\Delta_{ij})
\label{parameter}
\end{equation}

The solution of equation (\ref{Optic2}) is given by

\begin{equation}
\rho_{\tilde{\nu}} = \frac 12
\left(
\begin{matrix}
1 + \cos 2 \tilde{\theta}_{ij} e^{-\kappa t} & \sin^2\tilde{2\theta}_{ij}  e^{i 2\Delta_{ij} t} e^{-\kappa t/2} \\
\sin^2\tilde{2\theta}_{ij} e^{-i 2\Delta_{ij} t} e^{-\kappa t/2} &  1 - \cos 2 \tilde{\theta}_{ij} e^{-\kappa t}
\end{matrix}
\right)
\label{solution}
.
\end{equation}
From (\ref{solution}) it is easy to find the probability of the neutrino flavour oscillations

\begin{equation}
P_{\nu_e \to \nu_x} = \sin^2 2 \tilde{\theta}_{ij} \sin^2\left(\Delta_{ij} x\right) e^{-\kappa x/2} +
\frac 1 2 \left( 1 - \sin^2 2 \tilde{\theta}_{ij} e^{-\kappa x/2} - \cos^22 \tilde{\theta}_{ij}e^{-\kappa x}\right).
\label{Probability}
\end{equation}	
Here we consider the ultrarelativistic neutrinos and made the substitution $t \to x$.


In the case of an extreme relativistic electrons the decoherence parameter (\ref{parameter}) and the corresponding effect of quantum decoherence depend significantly upon the electron temperature $T$. This implies the need of the external environment high temperatures to make the influence of quantum decoherence on the neutrino evolution to be observable. Sufficiently high temperatures arise during supernovae bursts, where the electron and photon temperatures can reach values up to 30 MeV and 100 MeV, respectively \cite{Bolling_Janka_Lohs}. Therefore, we consider the proposed mechanism of quantum decoherence in supernovae environments. The electron matter potential $\sqrt{2} G_F n_e$ in supernovae is considered to be of the order $10^{-7}$ eV. For simplicity we consider an isotropic environment and neglect possible effects of collective neutrino oscillations.

The developed formalism can be applied both to the active-active and active-sterile neutrino oscillations. From equations (\ref{parameter}) one can see that the effect of the quantum decoherence strongly depends on the MSW effect and the temperature of the external environment. For the case of active-sterile neutrino oscillations (the case of active-active neutrino oscillations can be considered in the same way) the MSW effect occurs much closer  to the center of the SN \cite{Tamborra_Janka} explosion where high temperatures are expected.

The maximal value of neutrino decoherence parameter is of order $6*10^{-12} eV$, that is equivalent to the neutrino decoherence length $L_{dec} = \frac 2 \gamma \approx 60$ km$^{-1}$. It means that there is a significant suppression of neutrino oscillations at the distances of 60 km.

We have developed the method based on the theory of quantum electrodynamics of open systems for description of the quantum neutrino decoherence of the neutrino mass states. This method allows one to study the quantum decoherence  as a consequence of the radiative neutrino decay. We have considered the neutrino evolution propagating in the electron matter with the electromagnetic field and calculated the corresponding transition probability $P_{\nu_e \to \nu_\mu}$.  The exact analytical expression for the neutrino decoherence parameter and its energy dependence have been obtained. It has been also shown that the effect of neutrino decoherence can significantly modify the neutrino evolution in an extreme astrophysical environment peculiar to supernovae.

\end{document}